\documentclass[conference]{IEEEtran}

\RequirePackage[]{is2-2026}
\date{}
\title{Ananas: A Networked Audio System for Distributed Spatial Audio Applications}
\hypersetup{
 pdfauthor={Thomas Rushton},
 pdftitle={Ananas: A Networked Audio System for Distributed Spatial Audio Applications},
 pdfkeywords={},
 pdfsubject={},
 pdfcreator={Emacs 30.2 (Org mode 9.7.11)}, 
 pdflang={English}}
\begin{document}

\maketitle
\begin{abstract}
Spatial audio has an accessibility problem. Large-scale spatial and
immersive audio installations remain costly and exclusive, with high
barriers to entry. We propose ``Ananas,'' a networked audio system
consisting of network protocols, control software and microcontroller
firmware, with the aim of rendering spatial audio applications
accessible to a broad cohort of potential users. Based on a network of
low-cost microcontroller development boards programmed in the Faust
programming language, synchronised to nanosecond order via IEEE 1588
Precision Time Protocol, and controlled by approachable, digital audio
workstation-based tooling, Ananas represents an opportunity to disrupt
and democratise the spatial audio landscape. In this paper we
demonstrate Ananas's protocols for audio and control data
transmission and device discovery, its methods for media clock and
presentation time recovery, and its applications to wave field
synthesis and higher order ambisonics in a modular, scalable form.
\end{abstract}
\section{Introduction}
\label{sec:org04f01d4}

Despite significant interest in the research and commercial spheres,
access to large-scale spatial audio systems remains stubbornly out of
reach for many interested parties. Highly multichannel audio
interfaces are prohibitively expensive for all but the institutions
with the deepest pockets; implementing a spatial audio system
represents an investment in terms of money and expertise that places
the endeavour beyond the reach of many artists, researchers and other
potential users. In this paper we present Ananas --- \emph{``Ananas
necessitates another networked audio system''} --- a networked audio
system which, being based on a low-cost microcontroller development
board, open source software, and familiar digital audio workstation
(DAW) tools, seeks to lower the barrier to entry to spatial audio
techniques such as wave field synthesis
(WFS)~\cite{berkhoutAcoustic1993} and higher order ambisonics
(HOA)~\cite{zotterAmbisonics2019}.

The Ananas protocol suite describes a system of protocols for audio
and control data transmission, device discovery and
synchronisation. Taking advantage of multicast user datagram protocol
(UDP) groups for audio transmission and certain elements of control
data, Ananas minimises bandwidth, and, being based on IEEE 1588
Precision Time Protocol (PTP), ensures synchronisation amongst
distributed hardware modules. To our knowledge, Ananas is the first
open source system providing nanosecond order inter-module
synchronisation, following millisecond order synchronised systems such
as by Belloch et al.~\cite{bellochPerformance2021}.
\section{Background}
\label{sec:org647564c}

A wide array of networked audio protocols exist; why, then, create a
new one? The answer to this question lies in the time-sensitive nature
of spatial audio reproduction. In a distributed setting,
synchronisation of hardware nodes is critical, and existing protocols
are either proprietary, or do not cater to time-sensitive
applications.

Chief amongst proprietary systems is Dante~\cite{danteWhat2022}
(Digital Audio Network Through Ethernet), which uses PTP as its
time-exchange mechanism. Dante-enabled hardware is typically very
expensive, and, being a closed-source initiative, it is not suitable
for further scholarly enquiry. Another closed-source platform can be
found in the shape of the networking component of Elk Audio OS~\cite{turchetElk2021}.

A variety of open source protocols and systems are available, such as
JackTrip~\cite{caceresJackTrip2010a,caceresJackTrip2010},
NetJack~\cite{carotNetjack2009}, part of the JACK audio connection
kit, message-based systems such as Audio Over
OSC~\cite{ritschMessage2014} (the audio engine behind the audio
streaming application Sonobus\footnote{\url{https://sonobus.net}}), jamming-focused platforms such
as Jamulus~\cite{fischerCase2015a} and
Soundjack~\cite{renaudNetworked2007}, and the audiovisual performance
streaming platform LOLA~\cite{drioliNetworked2013}. What these
systems have in common, however, is a focus on networked music
performance and concert streaming, applications where low latency, and
not synchronicity, is of greatest concern. For time-sensitive
applications, this points to the need for a new networked audio
system, one that ensures synchronicity between distributed signal
processing modules.

Unlike Devonport and Foss~\cite{devonportDistribution2019}, and
Schultz et al.~\cite{schulzSeamLess2026} we opted not to use Audio
Video Bridging (AVB) or AES67 due to the high complexity associated
with implementing these protocols in a microcontroller context.
\subsection{The Hardware Platform}
\label{sec:org1d5bf6b}

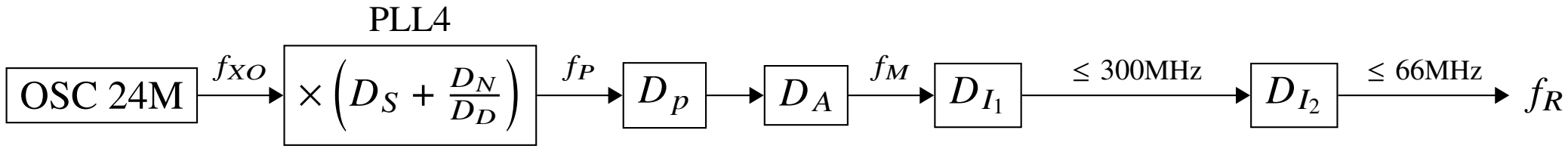
\begin{figure*}
\centering
\begin{tikzpicture}
  \node [draw] (osc) {OSC 24M};
  \node [draw] (pll4) [right=.75cm of osc] {\(\times \left(D_{S} + \frac{D_{N}}{D_{D}}\right)\)};
  \node (pll4label) [above=0cm of pll4] {\small{PLL4}};
  \node [draw] (dp) [right=.75cm of pll4] {\(D_{p}\)};
  \node [draw] (dA) [right=.5cm of dp] {\(D_{A}\)};
  \node [draw] (dI1) [right=.75cm of dA] {\(D_{I_{1}}\)};
  \node [draw] (dI2) [right=2cm of dI1] {\(D_{I_{2}}\)};
  \node (fR) [right=1.5cm of dI2] {\(f_{R}\)};

  \draw [-{Triangle}] (osc.east) -- (pll4.west) node [above, midway] {\scriptsize{\(f_{XO}\)}};
  \draw [-{Triangle}] (pll4.east) -- (dp.west) node [above, midway] {\scriptsize{\(f_{P}\)}};
  \draw [-{Triangle}] (dp.east) -- (dA.west);
  \draw [-{Triangle}] (dA.east) -- (dI1.west) node [above, midway] {\scriptsize{\(f_{M}\)}};
  \draw [-{Triangle}] (dI1.east) -- (dI2.west) node [above, midway] {\scriptsize{\(\leq 300 \text{MHz}\)}};
  \draw [-{Triangle}] (dI2.east) -- (fR.west) node [above, midway] {\scriptsize{\(\leq 66 \text{MHz}\)}};
\end{tikzpicture}
\caption{\label{fig:imxrt1062-clock}iMXRT1062 audio PLL and SAI1 clock divider registers.}
\end{figure*}

Effective deployment of a distributed spatial audio system is
dependent on the identification of a suitable hardware platform for
the distributed nodes. Such a platform should be low-cost and
powerful, with support for audio and ethernet. The Teensy 4.1
microcontroller development board~\cite{michonReal2019} was
identified as satisfying these requirements, costing less than € 50
with its audio and ethernet add-ons, and sporting a \qty{600}{\MHz}
ARM Core M7 CPU. Unlike other comparable devices, such as the
Raspberry Pi, its microcontroller, an NXP iMXRT1062 MCU, features a
dedicated audio subsystem, with high-resolution clock divider
registers for its audio phase locked loop (PLL); 30 bits are available
for the PLL's numerator and denominator registers, meaning that
nanosecond-order modifications of the audio sampling rate are
possible~\cite{nxpsemiconductorsIMX2021}.
\section{Implementation}
\label{sec:orga856373}

By convention, the output frequency produced by the iMXRT1062's
synchronous audio interface \texttt{SAI1\_CLK\_ROOT}, \(f_{R}\), must be 256 times
oversampled with respect to the target audio sampling rate, \(f_{s}\):

\begin{equation}
\label{eq:sai1-1}
f_{R} = 2^{8}f_{s}\;.
\end{equation}

The NXP MCUXpresso Config Tools
software~\cite{nxpsemiconductorsMCUXpresso2018} indicates that \(f_{R}\)
must be lower than or equal to \qty{66}{\MHz}, therefore

\begin{equation}
\label{eq:max-fs}
f_{s} \leq 257\,812.5\;\text{Hz}\;.
\end{equation}
\(f_{R}\) is found by dividing the frequency of \texttt{PLL4\_MAIN\_CLK},
\(f_{M}\), by the \texttt{SAI1\_CLK\_PRED}, \(D_{I_{1}}\), and \texttt{SAI1\_CLK\_PODF},
\(D_{I_{2}}\) fields of the \texttt{CCM\_CS1CDR} register:

\begin{equation}
\label{eq:sai1-2}
f_{R} = \frac{f_{M}}{D_{I_{1}}D_{I_{2}}}\;.
\end{equation}
\(f_{M}\) is found by dividing \(f_{P}\), the frequency of PLL4, by
the \texttt{POST\_DIV\_SELECT}, \(D_{p}\), and \texttt{AUDIO\_DIV}, \(D_{A}\), registers:

\begin{equation}
\label{eq:pll4-main}
f_{M} = \frac{f_{P}}{D_{p}D_{A}}\;.
\end{equation}
\(f_{P}\) is the result of multiplying the iMXRT1062's \qty{24}{\MHz}
clock source, \(f_{XO}\), by the PLL4 divider, \(D_{P}\):

\begin{equation}
\label{eq:pll4}
f_{P} = f_{XO}D_{P}\;,
\end{equation}
and \(D_{P}\) is, in turn, the sum of the \emph{PLL loop divider}, the
\texttt{DIV\_SELECT} field of the \texttt{CCM\_ANALOG\_PLL\_AUDIO} register, \(D_{S}\), with
the quotient of \texttt{CCM\_ANALOG\_PLL\_AUDIO\_NUM}, \(D_{N}\), the \emph{PLL
fractional loop divider} numerator register and
\texttt{CCM\_ANALOG\_PLL\_AUDIO\_DENOM}, \(D_{D}\), the denominator register:

\begin{equation}
\label{eq:pll4-divider}
D_{P} = D_{S} + \frac{D_{N}}{D_{D}}\;.
\end{equation}
See \cref{fig:imxrt1062-clock} for a diagrammatic representation of the
effect of the audio PLL and clock divider registers on the
\qty{24}{\MHz} clock source.

The \texttt{DIV\_SELECT} field has a valid range of \(27 \leq D_{S} \leq 54\),
and \(|D_{N}| < D_{D}\), so \(D_{P}\) describes (in theory) a number in the
following range:

\begin{equation}
\label{eq:divider-range}
\left(26 + \frac{1}{D_{D}}\right) \leq D_{P} \leq \left(54 + \frac{D_{D}-1}{D_{D}}\right)\;,
\end{equation}
with \(1 \leq D_{D} \leq 2^{30}\); effectively, \(26 < D_{P} <
55\). MCUXpresso Config Tools, and the iMXRT1062 reference
manual~\cite[§13.3.2.2]{nxpsemiconductorsIMX2021}, indicate,
however, that the output frequency range for PLL4 is:

\begin{equation}
\label{eq:pll4-range}
650\,\text{MHz} \leq f_{P} \leq 1\,300\,\text{MHz}\;,
\end{equation}
plus Config Tools contradicts the reference manual and states that
\(D_{N}\) must be a \emph{positive} 30 bit integer, so:

\begin{equation}
\label{eq:true-divider-range}
\left(27 + \frac{1}{12}\right) \leq D_{P} \leq \left(54 + \frac{1}{6}\right).
\end{equation}
Setting extreme (but valid) values for the divisors, we can find the
minimal valid \(f_{s}\):

\begin{equation}
\label{eq:fs-range}
309.9\;\text{Hz} \leq f_{s} \leq 257\,812.5\;\text{Hz}\;.
\end{equation}
\subsection{A General Expression for the Sampling Rate}
\label{sec:org621e1ac}

Bringing all of the above together, the general expression for
achieving a target sampling rate \(\hat{f}_{s}\) is

\begin{align}
\label{eq:general-fs}
\hat{f}_{s} &= \frac{f_{R}}{2^{8}} \nonumber \\
  &= \frac{f_{M}}{2^{8}D_{I_{1}}D_{I_{2}}} \nonumber \\
  &= \frac{f_{XO}D_{P}}{2^{8}D_{p}D_{A}D_{I_{1}}D_{I_{2}}} \nonumber \\
  &= \frac{24\times10^{6}}{2^{8}}\frac{D_{S}+\frac{D_{N}}{D_{D}}}{D_{p}D_{A}D_{I_{1}}D_{I_{2}}}\;.
\end{align}
This reckoning alone does not take into account certain constraints on
the system, however, such as \cref{eq:pll4-range} and the frequency at
the input of \(D_{I_{2}}\) being less than \qty{300}{\MHz} (see
\cref{fig:imxrt1062-clock}). There are too many free variables to solve
for any one of them, so an iterative approach to finding appropriate
register and field values is necessary.
\subsection{Setting the Sampling Rate}
\label{sec:setting-the-sampling-rate}
Say we wish to set a sampling rate of \(f_{s} = 48\;\text{kHz}\); our
target \(f_{R}\), \(\hat{f}_{R}\) equals, as per \cref{eq:sai1-1}

\begin{equation}
\label{eq:sai1-clk-root}
\hat{f}_{R} = 48\,000 \times 2^{8} = 12\,288\,000\;\text{Hz}\;.
\end{equation}
Strategically leaving \(D_{p}\) and \(D_{A}\) set to their default values of
1, values for \(D_{S}\), \(D_{N}\), \(D_{D}\), \(D_{I_{1}}\) and \(D_{I_{2}}\) can be
found:

\begin{itemize}
\item \(D_{S} = 29\)
\item \(D_{N} = 184\,000\,000\)
\item \(D_{D} = 1\,000\,000\,000\)
\item \(D_{I_{1}} = 3\)
\item \(D_{I_{2}} = 19\)
\end{itemize}
\subsection{Fine-Tuning the Sampling Rate}
\label{sec:orgfa7331e}

Say a set of divider register values has been established and some
manner of fine adjustment is desired, e.g., to compensate for clock
drift. \Cref{eq:general-fs} can be rewritten to find an expression for
an unknown \(D_{N}\):

\begin{equation}
\label{eq:unknown-dn}
D_{N} = D_{D}\left(
  \frac{2^{8}\hat{f}_{s}D_{p}D_{A}D_{I_{1}}D_{I_{2}}}{24 \times 10^{6}} - D_{S}
\right)\;.
\end{equation}
For very large \(D_{D}\) it may be advantageous to divide by the main
oscillator's frequency in \unit{\MHz} instead of \unit{\Hz}; to do so,
multiply the contents of the parenthesis in \cref{eq:unknown-dn} by one
million and divide the outside by one million to compensate:

\begin{equation}
\label{eq:unknown-dn-2}
D_{N} = \frac{D_{D}}{1 \times 10^{6}}\left(
  \frac{2^{8}\hat{f}_{s}D_{p}D_{A}D_{I_{1}}D_{I_{2}}}{24} - 1 \times 10^{6}D_{S}
\right)\;.
\end{equation}
\subsection{Media Clock Recovery}
\label{sec:media-clock}
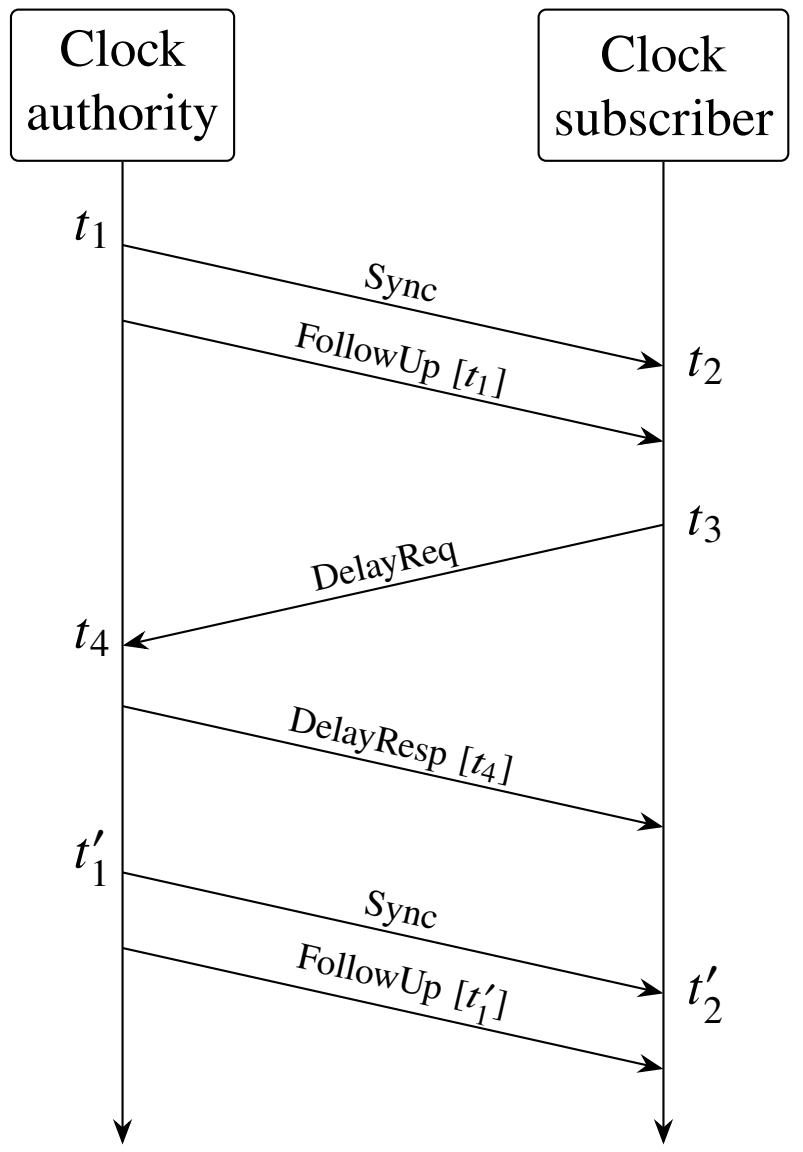
\begin{figure}[t]
\centering
\begin{tikzpicture}
  \pic{ptp};
\end{tikzpicture}
\caption{\label{fig:ptp}PTP packet flow.}
\end{figure}

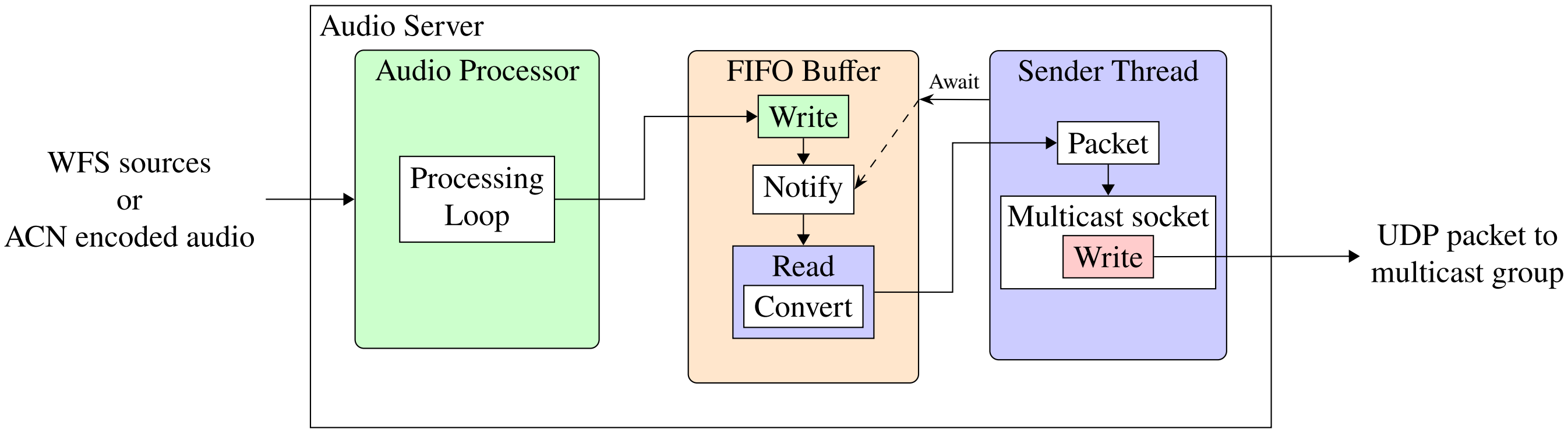
\begin{figure*}
\centering
\begin{tikzpicture}
\pic{server};
\end{tikzpicture}
\caption{\label{fig:server}Structural overview of the Ananas audio server.}
\end{figure*}

The Teensy's NXP iMXRT1062 MCU also features a dedicated hardware PTP
timer~\cite{schleusnerSubMicrosecond2024} thus allowing for effective
media clock recovery via an appropriate algorithm.

PTP is a time exchange mechanism
consisting of the exchange of multicast network packets between a
clock authority and a clock subscriber~\cite{leeProceedings2005,schleusnerSubMicrosecond2024}. PTP packet flow is illustrated in
\cref{fig:ptp}. Flow begins with the clock authority sending a \emph{sync}
packet; the subscriber notes the time, \(t_{2}\), at which this packet
arrives. The authority then sends a \emph{follow-up} packet which is
timestamped, \(t_{1}\), with the time of transmission. With \(t_{1}\)
and \(t_{2}\) in hand, the subscriber sends a \emph{delay request} packet
and notes the time, \(t_{3}\) of transmission; the authority notes the
time, \(t_{4}\) of reception of this packet and sends it to the
subscriber in a final \emph{delay response} packet. The above process then
repeats for the next time period, typically a second, beginning with
the receipt time \(t'_{2}\)of the following sync packet and the
receipt of the next follow-up packet containing timestamp
\(t'_{1}\). Once the subscriber has the above timestamp information it
can begin to compute its drift, \(\alpha\), delay, \(\delta\), and
offset, \(\theta\) relative to the authority:

\begin{align}
\label{eq:ptp1}
\alpha &= 1 - \frac{t'_{2} - t_{2}}{t'_{1} - t_{1}}\;, \\
\delta &= \frac{(t_{4} - t_{1}) - (t_{3} - t_{2})}{2}\;, \\
\theta &= t_{2} - t_{1} - \delta\;.
\end{align}

Armed with the above figures, the subscriber can feed them into a
proportional integral (PI) controller to yield a number of nanoseconds
for second \(n\) by which it should adjust its internal clock,

\begin{equation}
\label{eq:ptp-pi}
a_{n} = \alpha - K_{P}\theta_{n} + K_{I}\sum_{t=0}^{n}\theta\;,
\end{equation}
where \(K_{P}\) and \(K_{I}\) are co-efficients to be applied to the
proportional and integral terms respectively; small co-efficients,
typically \(K_{I} < K_{P} < 1\). The adjustment value \(a_{n}\) can be used to
set, for example, the per-cycle increment of a hardware ethernet timer
register.

Time exchange at the level of the ethernet subsystems of distributed
modules does not directly help us achieve audio sampling rate
parity. The time exchange information derived from PTP must be used to
condition the audio clocks of the modules such that sampling rate
parity is achieved; this is a process referred to as \emph{media clock
recovery}~\cite{crankshawMedia2021}.

In the case of the iMXRT1062 MCU this process is achieved by using the
nanoseconds-per-second PTP-derived adjustment value at second \(n\),
\(a_{n}\), to adjust the sampling rate. For a nominal sampling rate
\(f_{s}\) (e.g., \qty{48}{\kHz}), the sampling rate, \(\hat{f}_{s,n}\) that
``matches'' that of the time authority is found simply as

\begin{equation}
\label{eq:sampling-rate-recovery-1}
\hat{f}_{s,n} = f_{s}(1 + a_{n})\;.
\end{equation}
For example, if \(a_{n} = -1750\;\text{ns}\), we have

\begin{align*}
\label{eq:adjust-example}
\hat{f}_{s,n} &= f_{s}(1 - 1750 \times 10^{-9}) \\
  &= 48\,000 \times 0.999\,998\,25 \\
  &= 47\,999.916\;\text{Hz}\;.
\end{align*}
Inserting this value into \cref{eq:general-fs}, the expression for the
audio sampling rate of the iMXRT1062, yields

\begin{equation}
\label{eq:adjusted-fs}
47\,999.916 = \frac{24 \times 10^{6}}{2^{8}}\frac{D_{S} + \frac{D_{N,n}}{D_{D}}}{D_{p}D_{A}D_{I_{1}}D_{I_{2}}}\;,
\end{equation}
and rearranging, using \cref{eq:unknown-dn-2} to find the appropriate
value for the PLL4 numerator register

\begin{equation}
\label{eq:adjusted-dn}
\begin{split}
D_{N,n} = \frac{D_{D}}{1 \times 10^{6}}&\left(
  \frac{2^{8} \times 47\,999.916 \times D_{p}D_{A}D_{I_{1}}D_{I_{2}}}{24} - \right.\\
  &\left.1 \times 10^{6} D_{S}
\right)\;.
\end{split}
\end{equation}
Using the values derived in \cref{sec:setting-the-sampling-rate} yields

\begin{align*}
\label{eq:adjusted-dn-2}
D_{N,n} &= 1 \times 10^{3}\left(
  \frac{2^{8} \times 47\,999.916 \times 3 \times 19}{24} -
  29 \times 10^{6}
\right) \\
  &= 183\,948\,928\;,
\end{align*}
which can be used as a new value for the \texttt{CCM\_ANALOG\_PLL\_AUDIO\_NUM}
register of the iMXRT1062 MCU.

In practice \cref{eq:sampling-rate-recovery-1} does not suffice,
however, and some residual parts per billion drift remains. The
proposed solution is as follows: first, observing that at a nominal
sampling rate of \qty{48}{\kHz} for power of two audio buffer sizes
less than or equal to 128 there will be an integer number, \(I\), of
audio interrupts per second, take a representative nanoseconds figure,
\(t_{S}\), on audio system startup; second, every \(I\) interrupts,
reading the system's PTP timer, taking the nanoseconds part,
\(t_{\text{PTP}}\) and comparing with \(t_{S}\):

\begin{equation}
\label{eq:audio-ptp-offset}
\Delta_{n} = t_{\text{PTP}} - t_{S}\;;
\end{equation}
finally, applying another PI controller:

\begin{equation}
\label{eq:sampling-rate-recovery-2}
\hat{f}_{s,n} = f_{s}\left(
  1 + a_{n} + K_{P_{\Delta}}\Delta_{n} + K_{I_{\Delta}}\sum_{t=0}^{n}\Delta
\right)\;.
\end{equation}
These additional correction factors ensure reliable media clock
recovery indefinitely.
\section{The Ananas Networked Audio System}
\label{sec:orgc21c928}

Ananas is a system for time-sensitive networked audio for supporting
distributed approaches to sound field synthesis. In the sections that
follow we describe the various components of the Ananas system.
\subsection{The Audio Server}
\label{sec:org54217c4}

The Ananas audio server is a C++ library written using the JUCE
framework.\footnote{JUCE v8.0.6 \url{https://juce.com}.} On startup, the library initialises a series of
threads for handling audio transmission, reading PTP timestamps and
listening for connected clients, amongst other tasks. The audio
transmission part of the server is illustrated in \cref{fig:server}. The
server receives a collection of input channels, either monophonic WFS
sources or ambisonics channel numbering (ACN) encoded audio channels,
and writes these to an intermediate first-in-first-out (FIFO)
buffer. When the FIFO contains enough samples to fill an outgoing
audio packet, the sender thread is notified and reads from the FIFO to
fill a packet for transmission. The sender thread delivers the packet
to a UDP multicast socket and transmits it to the network.
\subsection{The Ananas Audio Protocol}
\label{sec:org241990b}

\begin{figure}[t]
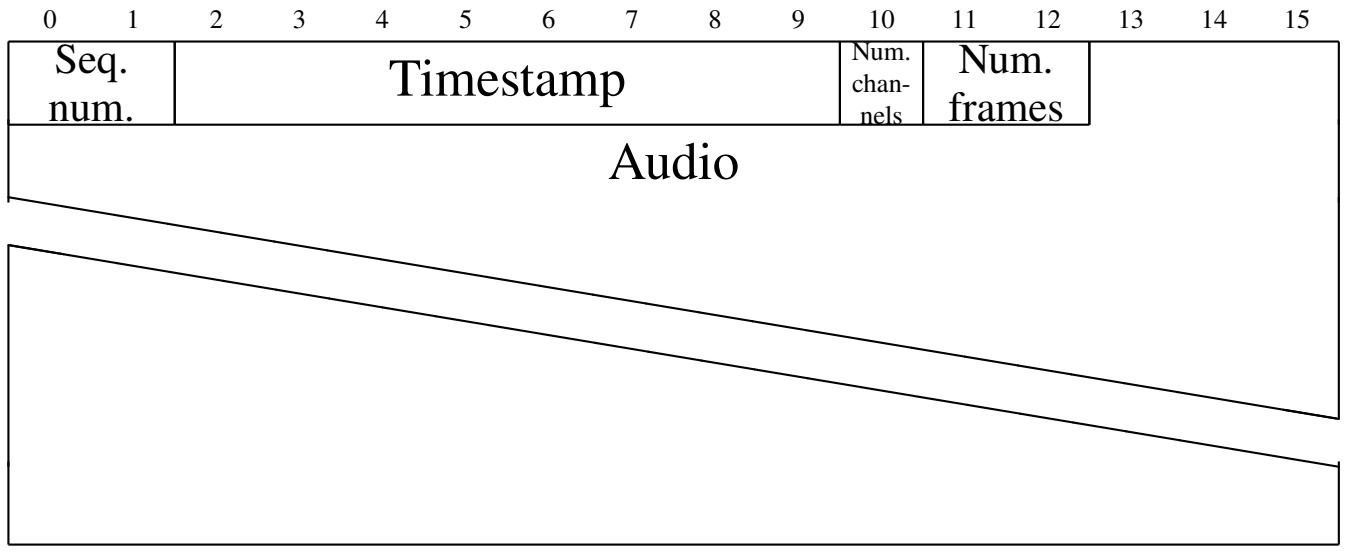

\centering
\begin{bytefield}[bitwidth=.55cm,bitheight=.55cm]{16}
  \bitheader{0-15} \\
  \bitbox{2}{\footnotesize{Seq. num.}} & \bitbox{8}{Timestamp} & \bitbox{1}{\tiny{Num. channels}} & \bitbox{2}{\footnotesize{Num. frames}} & \bitbox[lrt]{3}{} \\
  \wordbox[lr]{1}{Audio} \\
  \skippedwords \\
  \wordbox[lrb]{1}{} \\
\end{bytefield}
\caption{\label{fig:audio-packet}Structure of an Ananas audio packet.}
\end{figure}

Being first and foremost concerned with the transmission of audio
data, at the heart of the Ananas system is the Ananas Audio Protocol
(AAP). A UDP multicast-based protocol, using address \texttt{224.4.224.4},
taken from the second ad-hoc block of IANA's multicast address
space~\cite{meyerIANA2010}, AAP consists of a stream of sequence
numbered, timestamped packets. The structure of an AAP packet is given
in \cref{fig:audio-packet}. It was found that, particularly under
high-bandwidth conditions, packets sporadically arrive out of order,
hence the inclusion of the sequence number, which is used at the
receiving side to ensure the integrity of the packet stream. For
timestamping, the server reads a PTP follow-up packet, adds a small
reproduction offset (for packets of 16 frames it was found that an
offset of \(1/70\) s was most appropriate) and writes the timestamp to
the outgoing packet; successive packets are then updated with the
expected inter-packet interval, e.g., for 16 frames at \qty{48}{\kHz}
this is \(333\,333\frac{1}{3}\) ns.

Completing the packet header are the number of channels and number of
frames in the packet, used at the receiving side for decoding the
byte-encoded audio data that follows. At 16 frames per packet it is
possible to transmit up to 44 channels of 16-bit audio while
respecting ethernet's 1500 byte \emph{maximum transmissible unit}.

\newpage

\subsection{The Ananas Client Description Protocol}
\label{sec:org236f221}

\begin{figure}
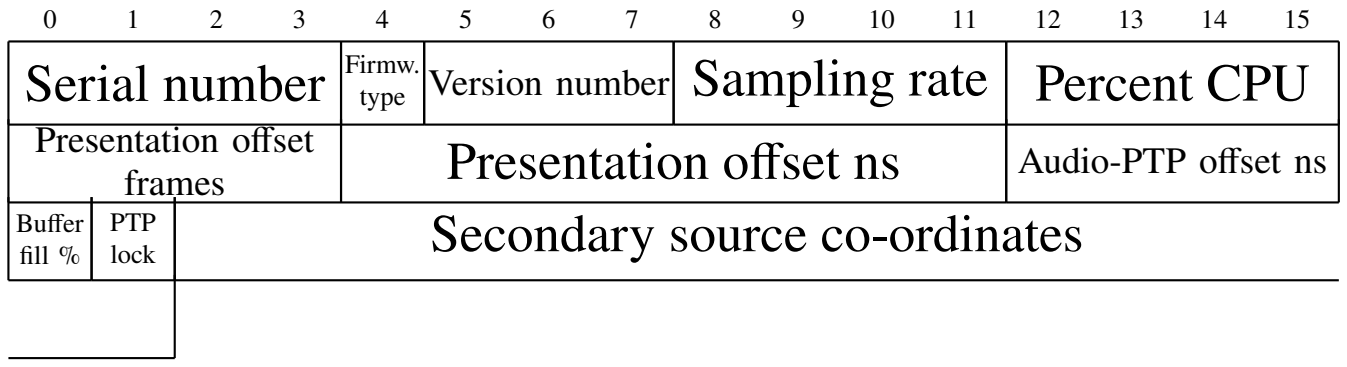

\centering
\begin{bytefield}[bitwidth=.55cm, bitheight=.55cm]{16}
      \bitheader{0-15} \\
      \bitbox{4}{Serial number} & \bitbox{1}{\tiny{Firmw. type}} & \bitbox{3}{\scriptsize{Version number}} & \bitbox{4}{Sampling rate} & \bitbox{4}{Percent CPU} \\
      \bitbox[lrb]{4}{\scriptsize{Presentation offset frames}} & \bitbox[lrb]{8}{Presentation offset ns} & \bitbox[lrb]{4}{\scriptsize{Audio-PTP offset ns}} \\
      \bitbox[lrb]{1}{\tiny{Buffer fill \%}} & \bitbox[lrb]{1}{\tiny{PTP lock}} & \bitbox[lb]{14}{Secondary source co-ordinates} \\
      \bitbox[rb]{2}{}
    \end{bytefield}
\caption{\label{fig:client-packet}Structure of an Ananas Client Description packet.}
\end{figure}

Clients announce their presence on the network with periodic UDP
multicast packets on address \texttt{224.4.224.6}. A client is deemed to be
connected if the server has received an Ananas Client Description
Protocol (ACDP) packet within the last second. Each packet contains a
variety of information about a given client, such as its serial
number, firmware type and version, reported sampling rate, buffer fill
percentage and PTP lock status, etc., information which can be
displayed in a digital audio workstation plugin interface. The
structure of an ACDP packet is given in \cref{fig:client-packet}.
\subsection{The Time Authority}
\label{sec:time-authority}
One Teensy 4.1 is designated as an \emph{authoritative source of time} for
the entire system. It is used as a USB audio device for the server
machine, a general purpose laptop computer, and as a PTP clock
authority for all other devices on the network. Since PTP time and USB
audio timing on the time authority are governed by the same crystal
oscillator, in serving this hybrid role the time authority ensures
synchronicity between the server machine, the ethernet switch, which
acts as a PTP boundary clock, and, by extension, all other Teensy
4.1-based modules in the system.
\subsection{Control Systems}
\label{sec:orgb6297ca}

Control data is sent as OSC messages over UDP. For WFS, hardware
modules are assigned their positions in the speaker array, i.e. given
their secondary source (loudspeaker) positions over unicast
UDP. E.g., a given module may receive its secondary source x-position
via an OSC message of the form \texttt{/ss/0/x 1.5}, indicating that its first
secondary source is at x-coordinate \qty{1.5}{\m}. WFS virtual sound
source positions are delivered as multicast OSC messages on address
\texttt{224.4.224.5} and are of the form \texttt{/vs/0/y -0.5}, indicating that the
first virtual sound source's y-coordinate should be \qty{-0.5}{\m}.

For ambisonics, modules instead receive their decoder coefficients
over unicast UDP, again as OSC messages. A module may for example
receive a message of the form \texttt{/dm/0/0 0.2385}, indicating that it
should use a coefficient of 0.2385 for the first element in the first
row of the decoder matrix.
\subsection{System Overview}
\label{sec:orgc54b03a}

\begin{figure}
\centering
\begin{tikzpicture}
\pic{systemOverview};
 \matrix [draw] at (3.75, 0) {
    \draw [ctrlArrow] (0,0) -- (.5,0); & \node [legend] {Control data (UDP)}; \\
    \draw [audio] (0,0) -- (.5,0); & \node [legend] {Audio (UDP)}; \\
    \draw [-{Triangle[open]}, usbColour] (0,0) -- (.5,0); & \node [legend] {USB audio}; \\
    \draw [-{Triangle}] (0,0) -- (.5,0); & \node [legend] {PTP}; \\
  };
\end{tikzpicture}
\caption{\label{fig:system-overview}Overview of the Ananas networked audio system.}
\end{figure}
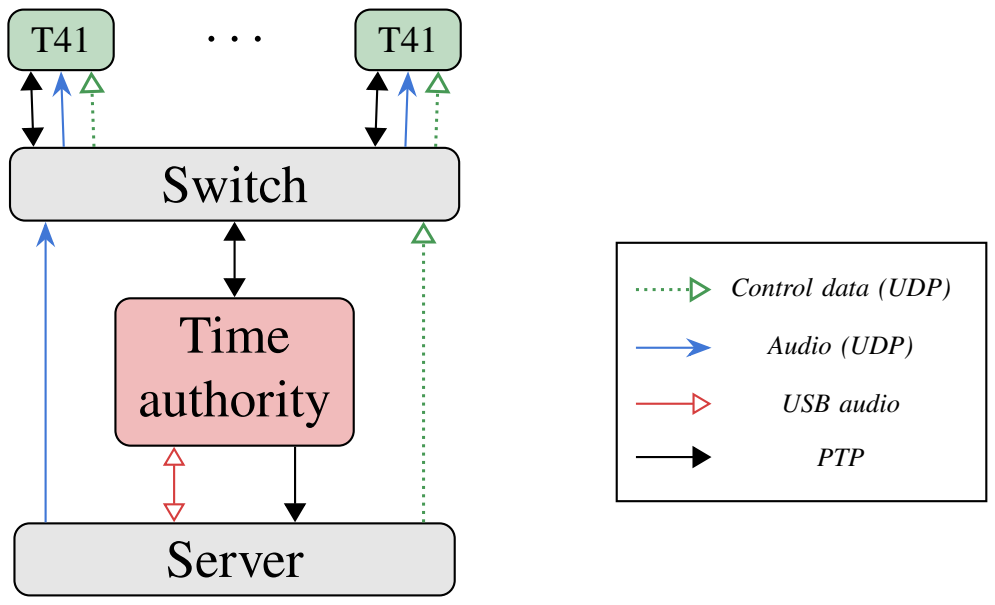

An overview of the Ananas networked audio system is given in
\cref{fig:system-overview}. The time authority governs USB audio timing
on the server machine and PTP time for the switch, a MikroTik
CRS326-24G-2S+IN\footnote{\url{https://mikrotik.com/product/crs326_24g_2s_in}} chosen for its high port count, PTP support and
low cost (\(\sim\)€ 200), and other connected modules. The server sends audio
and control data to the switch, which forwards this data on to the
connected modules, which perform their part of a distributed spatial
audio algorithm.

Code for the Ananas server can be found at
\url{https://github.com/hatchjaw/ananas-server}; code for the Ananas
client and time authority can be found at
\url{https://github.com/hatchjaw/ananas-client}.
\section{Deployment}
\label{sec:org2ab376e}

\begin{figure}[htbp]
\centering
\includegraphics[width=.9\linewidth]{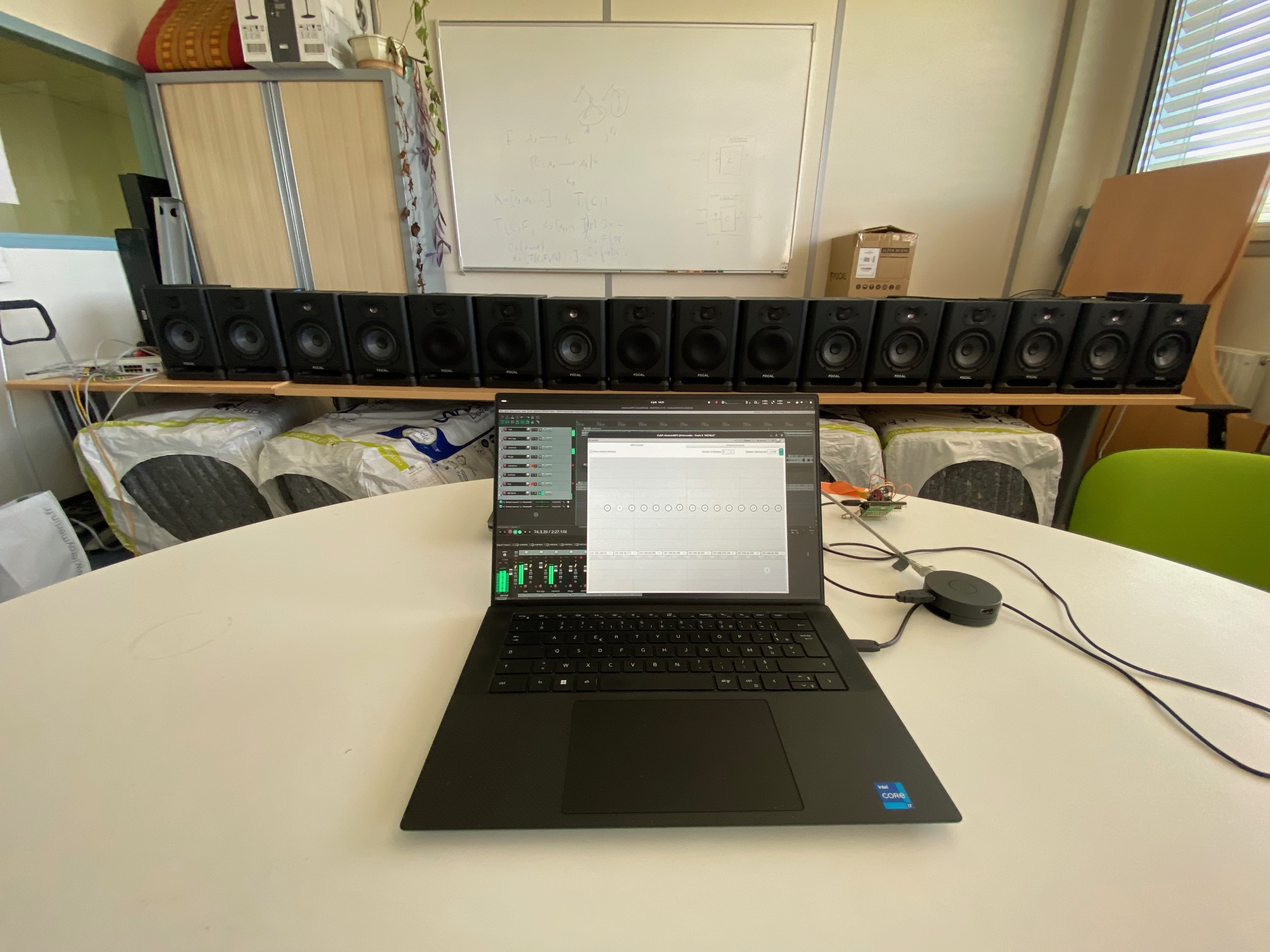}
\caption{\label{fig:ananas-wfs}Ananas driving a 16-channel WFS system.}
\end{figure}

\begin{figure*}
\centering
\includegraphics[width=.9\linewidth]{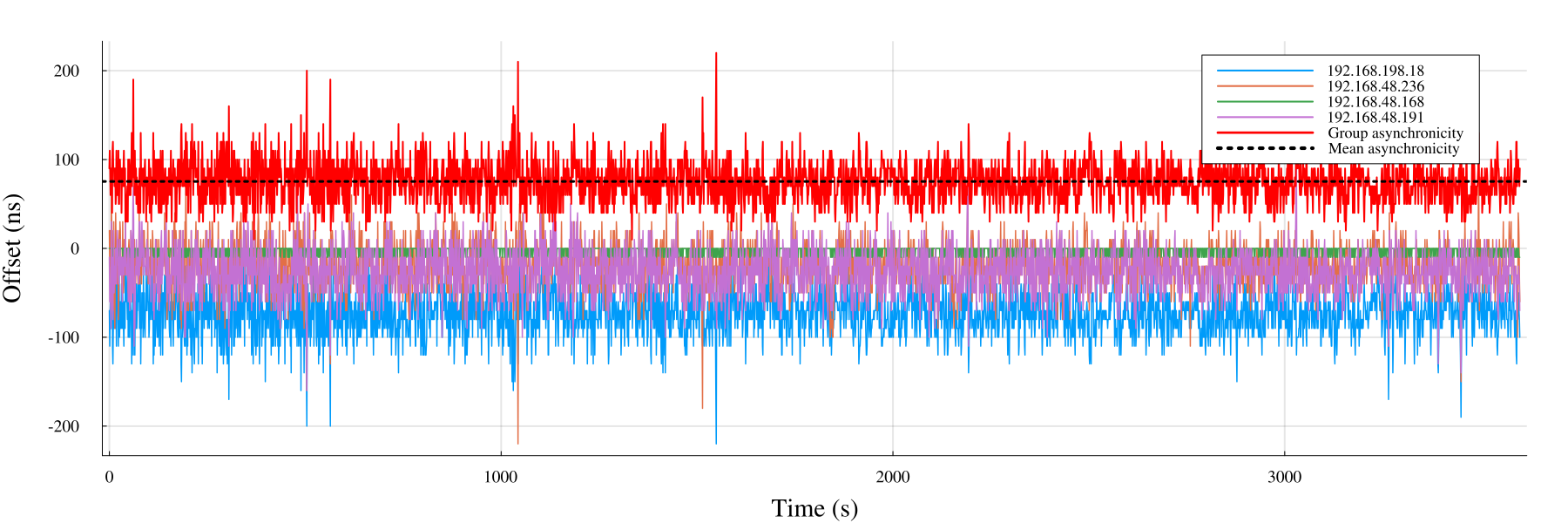}
\caption{\label{fig:synchro}Synchronisation measurements for a group of four Teensy modules reproducing a pulse-per-second audio signal. Group asynchronicity represents the instantaneous range of offsets for all four modules.}
\end{figure*}

\begin{figure}[htbp]
\centering
\includegraphics[width=.9\linewidth]{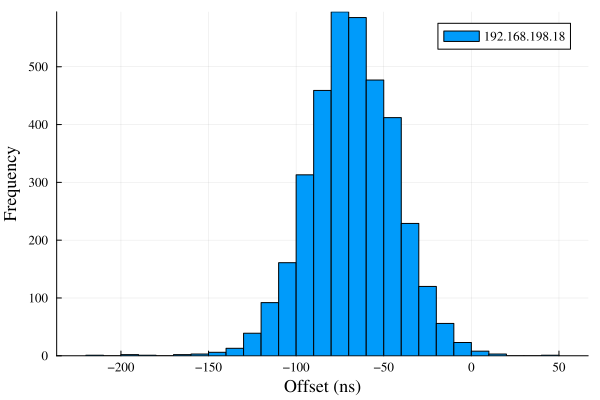}
\caption{\label{fig:histo}Histogram of synchronisation measurements for one of the clients in \cref{fig:synchro}.}
\end{figure}

Deployed as per \cref{fig:system-overview}, Ananas can operate either as
a WFS or a HOA system. DSP, programmed in the Faust language, is
distributed amongst the hardware modules; for WFS, modules need only
know the spatial coordinates of their loudspeakers and the
coordinates of the virtual sound sources to produce their part of the
synthesised sound field; for HOA, modules are provided with the rows
of the decoder matrix for which they are responsible and delivered
ACN-encoded audio streams.

An example WFS setup is pictured in \cref{fig:ananas-wfs}. Timing in the
system is governed by the time authority Teensy described in
\cref{sec:time-authority}. The server machine, a general purpose laptop
computer, runs the Reaper\footnote{\url{https://reaper.fm}} digital audio workstation with a CLAP
format WFS control plugin assigned to a 16-channel track, and sends
monophonic audio streams to a UDP multicast group managed by a single
switch, which forwards the data on to the hardware modules. Eight
modules are used in this setup for a total of sixteen output
channels. The server sending packets containing sixteen frames of
audio data, it is possible to transmit up to 44 channels, each
representing either a WFS virtual source or an ACN-format channel; up
to fifth order HOA is supported.

Thanks to the media clock recovery algorithm described in
\cref{sec:media-clock}, modules remain synchronised to within around
\qty{200}{\nano\second} at all times. \Cref{fig:synchro} shows an hour's
worth of synchronicity measurements for a group of four hardware
modules. The modules were connected to a Digilent Analog Discovery 2
logic analyser and their I\textsuperscript{2}S output measured in Digilent Waveforms
software. The server delivered a pulse-per-second audio signal and
offsets were measured at the rising edge of the pulse. The mean
asynchronicity of \(\sim\)\qty{80}{\nano\second} falls well within the AES
guideline for time-sensitive applications of 5\% of the reference frame
(\(\sim\)\qty{1}{\micro\second} at \qty{48}{\kilo\hertz}) and corresponds with a
propagation discrepancy in air of roughly \qty{27.4}{\micro\meter}. A
histogram of synchronisation data for client \texttt{192.168.198.18} is
presented in \cref{fig:histo}.
\section{Conclusion}
\label{sec:orgee88c54}

In this paper we have presented \emph{Ananas}, a new networked audio system
for time-sensitive applications. With its low-cost,
microcontroller-based hardware platform, it is modular and scalable
and familiar digital audio workstation tools make it approachable to
non-technical users. Modules are synchronised to extremely high
precision thanks to media clock and presentation time recovery
algorithms. Ananas represents an opportunity to democratise spatial
audio, rendering WFS and HOA accessible to a broad potential userbase.

In terms of future work, we aim to increase the channel count
available on each hardware module via time-division multiplexing,
further reducing the cost per output channel associated with the
Ananas system.

\bibliographystyle{IEEETran}
\bibliography{IS22026}
\end{document}